\documentclass[12pt]{article}

\usepackage{graphicx}
\begin{document}

\begin{center}
{\bf Asymptotic Reissner-Nordstr\"om solution within nonlinear electrodynamics } \\
\vspace{5mm} S. I. Kruglov
\footnote{E-mail: serguei.krouglov@utoronto.ca}
\underline{}
\vspace{3mm}

\textit{Department of Chemical and Physical Sciences, University of Toronto,\\
3359 Mississauga Road North, Mississauga, Ontario L5L 1C6, Canada} \\
\vspace{5mm}
\end{center}

\begin{abstract}
A model of nonlinear electrodynamics coupled with the gravitational field is studied.
We obtain the asymptotic black hole solutions at $r\rightarrow 0$ and $r\rightarrow \infty$. The asymptotic at $r\rightarrow 0$ is shown, and we find corrections to the Reissner-Nordstr\"om solution and Coulomb's law at $r\rightarrow\infty$. The mass of the black hole is evaluated having the electromagnetic origin. We investigate the thermodynamics of charged black holes and their thermal stability. The critical point corresponding to the second-order phase transition (where heat capacity diverges) is found. If the mass of the black hole is greater than the critical mass, the black hole becomes unstable.
\end{abstract}

\section{Introduction}

Nonlinear electrodynamics (NLED) can solve the problem of the singularity of an electric field in the origin of charged pointlike particles as well as the problem of infinite electromagnetic energy. For weak fields NLED may be converted into Maxwell's electrodynamics so that Maxwell's electrodynamics can be considered as an approximation. For strong electromagnetic fields classical electrodynamics has to be modified \cite{Jackson} because the self-interaction of photons is important.
Born and Infeld \cite{Born} (BI) proposed a model of NLED so that there is an upper limit on the electric field at the origin of charged particles and the total electric energy is finite. Thus, the BI model of NLED may solve the problem of singularity in the classical theory. The BI action also was obtained from the low-energy effective action of the superstring theory \cite{Fradkin}, \cite{Tseytlin}. In Maxwell's electrodynamics and in the BI NLED, the dual invariance holds \cite{Gibbons}, \cite{Gibbons1}. Within various models of NLED \cite{Shabad}, \cite{Kruglov}, \cite{Kruglov2}, the problems of singularities and the infinity of a total electromagnetic energy of charged particles can be solved.

In the early epoch of the Universe electromagnetic and gravitational fields were very strong, and therefore, nonlinear effects should be taken into account. In addition, nonlinear electromagnetic fields can drive the Universe acceleration \cite{Garcia}, \cite{Camara}.
In \cite{Elizalde}-\cite{Kruglov4}, the magnetic universe was considered and the stochastic magnetic fields result in inflation of the Universe. BI cosmology was investigated in \cite{Garcia}. It should be mentioned that electromagnetic fields in BI electrodynamics do not drive the Universe to accelerate \cite{Novello1} and, in addition, the BI model suffers causality problems \cite{Quiros}.
Recently, the exact black hole solutions in the framework of general relativity (GR) coupled with NLED were obtained in \cite{Ayon}-\cite{Kruglov6}.

In this paper, we obtain a solution for charged black holes in GR theory coupled with NLED proposed in \cite{Kruglov}. Corrections to the Coulomb law and to the Reissner-Nordstr\"om (RN) black hole solution are found. We also study the thermodynamics of black holes and phase transitions.

The paper is organized as follows. In Sec. II, we introduce a model of NLED and
study energy conditions. NLED coupled to gravitational field is investigated in Sec. III. Corrections to the Coulomb law are obtained. We investigate the P frame that corresponds to the electric-magnetic duality.
The mass of the black hole is calculated in Sec. IV. We find the black hole solution and obtain corrections to the RN solution. We show that the solution possesses the Reissner-Nordstr\"om  asymptotic. In Sec. V thermodynamics of black holes and phase transitions are investigated. It was demonstrated that there is no
first-order phase transition when the temperature of black holes changes the sign. We obtain the critical point corresponding to the second-order phase transition. Sec. VI is devoted to a conclusion.

We use the units $c=\hbar=1$ and the metric $\eta=\mbox{diag}(-1,1,1,1)$.

\section{Nonlinear electrodynamics and energy conditions}

Let us consider NLED with the Lagrangian density proposed in \cite{Kruglov}
\begin{equation}
{\cal L}_{em} = -\frac{{\cal F}}{2\beta{\cal F}+1}.
 \label{1}
\end{equation}
The parameter $\beta$ possesses the dimension of (length)$^4$, and gives the upper bound on the possible electric field strength \cite{Kruglov}, ${\cal F}=(1/4)F_{\mu\nu}F^{\mu\nu}$, and $F_{\mu\nu}$ is the field strength. The electric field of a pointlike charged particle has the finite value at the origin and the electromagnetic energy of charged particles is also finite.
The energy-momentum tensor is given by \cite{Kruglov}
\begin{equation}
T_{\mu\nu}=-\frac{1}{(2\beta{\cal F}+1)^2}\left[F_{\mu}^{~\alpha}F_{\nu\alpha}-g_{\mu\nu}{\cal F}(2\beta{\cal F}+1)\right],
 \label{2}
\end{equation}
and the trace of the energy-momentum tensor is nonzero,
\begin{equation}
{\cal T}\equiv T_{\mu}^{~\mu}=\frac{8\beta {\cal F}^2}{\left(2\beta{\cal F}+1\right)^2}.
\label{3}
\end{equation}
At $\beta\rightarrow 0$, one comes to Maxwell's electrodynamics with zero trace.
The scale and dilatation invariance are broken due to the presence of the dimensional parameter $\beta$.

The weak energy condition (WEC) \cite{Hawking} is defined as
\begin{equation}
\rho\geq 0,~~~\rho+p_m\geq 0 ~~(m=1,~2,~3),
\label{4}
\end{equation}
where $\rho$ is the energy density, and $p_m$ are principal pressures. Equations (4) guarantee that the energy density is non-negative for any local observer. For the case $\textbf{B}=0$, we find from Eq. (2),
\begin{equation}
\rho=T_0^{~0}=\frac{E^2(1+\beta E^2)}{2(1-\beta E^2)^2},
\label{5}
\end{equation}
\begin{equation}
p_m=-T_m^{~m}=\frac{E^2(1-\beta E^2)-2E_m^2}{2(1-\beta E^2)^2}~~~~(m=1,2,3),
\label{6}
\end{equation}
and there is not the summation in the index $m$ in Eq. (6), $E^2=E_1^2+E_2^2+E_3^2$. The plot of the function $\rho$ is given in Fig. 1.
\begin{figure}[h]
\includegraphics[height=3.0in,width=3.0in]{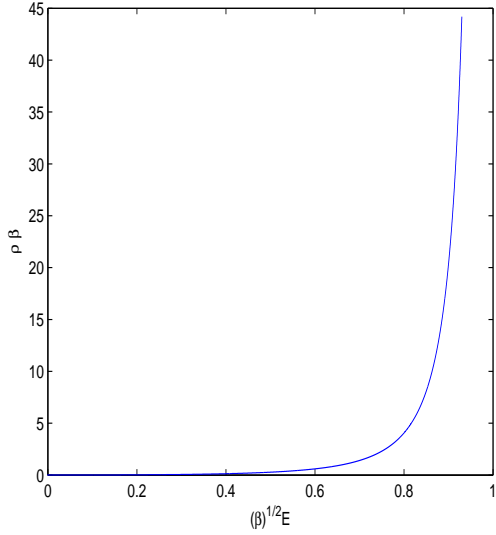}
\caption{\label{fig.1}The function  $\rho\beta$  vs $E\sqrt{\beta}$.}
\end{figure}
One can see from Eq. (5) that $\rho\geq 0$. By virtue of Eqs. (5),(6) for the case $\textbf{B}=0$, we obtain
\begin{equation}
\rho+p_1=\frac{E_2^2+E_3^2}{(1-\beta E^2)^2},~~
\rho+p_2=\frac{E_1^2+E_3^2}{(1-\beta E^2)^2},~~
\rho+p_3=\frac{E_1^2+E_2^2}{(1-\beta E^2)^2}.
\label{7}
\end{equation}
Thus, WEC is satisfied for any values of the electric field. It should be noted that $\beta E^2<1$ because the maximum electric field at the origin of charged particles is $E_{max}=1/\sqrt{\beta}$ \cite{Kruglov}.

The dominant energy condition (DEC) \cite{Hawking} reads
\begin{equation}
\rho\geq 0,~~~\rho+p_m\geq 0,~~~\rho-p_m\geq 0~~(m=1,~2,~3).
\label{8}
\end{equation}
From Eqs. (5) and (6) (at $\textbf{B}=0$), one finds
\begin{equation}
\rho-p_1=\frac{E_1^2+\beta E^4}{(1-\beta E^2)^2},~~
\rho-p_2=\frac{E_2^2+\beta E^4}{(1-\beta E^2)^2},~~
\rho-p_3=\frac{E_3^2+\beta E^4}{(1-\beta E^2)^2}.
\label{9}
\end{equation}
Equations (8) hold and, as a result, the speed of sound is less than the speed of light.

The strong energy condition (SEC) \cite{Hawking} requires
\begin{equation}
\rho+\sum_{m=1}^3p_m\geq 0.
\label{10}
\end{equation}
With the aid of Eqs. (5) and  us(6) for the case $\textbf{B}=0$, we obtain
\begin{equation}
\rho+\sum_{m=1}^3p_m=\frac{E^2}{1-\beta E^2},
\label{11}
\end{equation}
and SEC (10) is satisfied. The pressure at $\textbf{B}=0$ is
\[
p={\cal L}+\frac{E^2}{3}\frac{\partial {\cal L}}{\partial {\cal F}}=\frac{E^2(1-3\beta E^2)}{6(1-\beta E^2)^2}=\frac{1}{3}\sum_{m=1}^3p_m.
\]
Then SEC (10) can be represented as $\rho +3p\geq 0$ which tells us that the electric field can not drive the acceleration of the Universe. In the magnetic Universe, the acceleration of the Universe occurs in the model under consideration \cite{Kruglov3}.

\section{NLED and black holes}

The action of NLED coupled with GR is given by
\begin{equation}
S=\int d^4x\sqrt{-g}\left[\frac{1}{2\kappa}R+ {\cal L}\right].
\label{12}
\end{equation}
The $R$ is the Ricci scalar, $\kappa=8\pi G\equiv M_{Pl}^{-2}$, $M_{Pl}$ is the reduced Planck mass and $G$ is the  Newton constant. By varying action (12), one can find the Einstein equation and equations for NLED:
\begin{equation}
R_{\mu\nu}-\frac{1}{2}g_{\mu\nu}R=-\kappa T_{\mu\nu},
\label{13}
\end{equation}
\begin{equation}
\partial_\mu\left(\frac{\sqrt{-g}F^{\mu\nu}}{(1+2\beta {\cal F})^2}\right)=0.
\label{14}
\end{equation}
Our goal is to obtain the static charged black hole solutions to Eqs. (13),(14). We explore the spherically symmetric line element in
$(3+1)$-dimensional spacetime as follows:
\begin{equation}
ds^2=-f(r)dt^2+\frac{1}{f(r)}dr^2+r^2(d\vartheta^2+\sin^2\vartheta d\phi^2).
\label{15}
\end{equation}
Implying that the vector-potential possesses nonzero component $A_0(r)$, ${\cal F}=-[E(r)]^2/2$  ($\textbf{B}=0$), from Eq. (14)
we obtain
\begin{equation}
\partial_r\left(\frac{r^2 E(r)}{(1-\beta [E(r)]^2)^2}\right)=0.
\label{16}
\end{equation}
Integrating Eq. (16), one finds the equation
\begin{equation}
r^2E(r)=Q(1-\beta [E(r)]^2)^2,
\label{17}
\end{equation}
where $Q$ is the constant of integration. We introduce the dimensionless variables \footnote{We use the variables which are different from \cite{Kruglov5}.}
\begin{equation}
x=\frac{r}{\beta^{1/4} \sqrt{Q}},~~~~y=\beta^{1/4}\sqrt{E(r)}.
\label{18}
\end{equation}
Then Eq. (17) becomes
\begin{equation}
y^4+xy-1=0.
\label{19}
\end{equation}
The positive real solution to Eq. (19) is
\[
y=\sqrt{\frac{3^{1/4}x}{4\sqrt{\sinh(\varphi/3)}}-\frac{\sinh(\varphi/3)}{\sqrt{3}}}-\frac{\sqrt{\sinh(\varphi/3)}}{3^{1/4}},
\]
\begin{equation}
\sinh(\varphi)=\frac{3^{3/2}}{16}x^2.
\label{20}
\end{equation}
The plot of the function $y(x)$ is given by Fig. 2.
\begin{figure}[h]
\includegraphics[height=3.0in,width=3.0in]{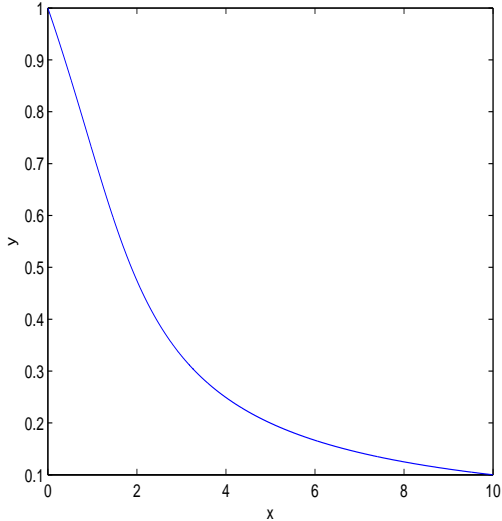}
\caption{\label{fig.2}The function $y$ vs $x$.}
\end{figure}
At $x\rightarrow 0$, we have $y\rightarrow 1$ and from Eq. (18) one obtains the finite value of the electric field in the center $E(0)=1/\sqrt{\beta}$ \cite{Kruglov}. Thus, we have the maximum of the electric field at the origin.

Let us investigate the functions $E(r)$, $A_0(r)$.
The asymptotic behavior of the function $y(x)$ (20) at $x\rightarrow\infty$ is given by
\begin{equation}
y=\frac{1}{x}+0.592592\frac{1}{x^{11/3}}-0.209879\frac{1}{x^{15/3}}+{\cal O}(x^{-19/3}).
\label{21}
\end{equation}
From Eqs. (18) and (21), we find at $r\rightarrow \infty$ the asymptotic value of the electric field:
\begin{equation}
E(r)=\frac{Q}{r^2}+1.185184\frac{Q^{7/3}\beta^{2/3}}{r^{14/3}}+{\cal O}(r^{-18/3}).
\label{22}
\end{equation}
Equation (22) shows that the $Q$ is the charge and corrections to the Coulomb law at $r\rightarrow\infty$ are in the order of $r^{-14/3}$. At $\beta=0$, one comes to the Coulomb law $E=Q/r^2$ of Maxwell's electrodynamics.
Integrating the function (22), we obtain the the asymptotic value of the electric potential ($A_0(r)=\int E(r)dr$) at $r\rightarrow\infty$:
\begin{equation}
A_0(r)=-\frac{Q}{r}-\frac{32Q^{7/3}\beta^{2/3}}{99r^{11/3}}+{\cal O}(r^{-15/3}).
\label{23}
\end{equation}
The Taylor series of $y(x)$ at $x\rightarrow 0$  ($r\rightarrow 0$) is given by
\begin{equation}
y(x)=1-0.25x-0.0312499x^2+0.00341799x^4+{\cal O}(x^5).
\label{24}
\end{equation}
From Eqs. (18) and (24), we obtain the asymptotic for the electric field at $r\rightarrow 0$:
\begin{equation}
E(r)=\frac{1}{\sqrt{\beta}}-\frac{r}{2\sqrt{Q}\beta^{3/4}}+\frac{r^2}{2\times 10^6 Q\beta}+{\cal O}(r^3).
\label{25}
\end{equation}
From Eq. (25), we find the expected result $E(0)=1/\sqrt{\beta}$. One can obtain from Eq. (25) the electric potential in the center.
As a result, the electric field and the electric potential are finite at the origin of the charged objects and there are no singularities. This behavior of the electric field at $r=0$ is due to non-Maxwellian character of NLED considered.
Such situation holds also in different nonlinear electrodynamics \cite{Born}, \cite{Kruglov2}.

\subsection{P framework}

By virtue of a Legendre transformation \cite{Garcia1}, we can arrive at another form of NLED. Let us consider the tensor $P_{\mu\nu}={\cal L}_{\cal F}F_{\mu\nu}/2$ (${\cal L}_{\cal F}=\partial{\cal L}/\partial{\cal F})$. Then with the aid of (1) one obtains the invariant
\begin{equation}
P=P_{\mu\nu}P^{\mu\nu}=\frac{{\cal F}}{(1+2\beta {\cal F})^4}.
\label{26}
\end{equation}
The Hamilton-like variable becomes
\begin{equation}
{\cal H}=2{\cal F}{\cal L}_{\cal F}-{\cal L}=\frac{{\cal F}(2\beta{\cal F}-1)}{(1+2\beta {\cal F})^2}.
\label{27}
\end{equation}
Comparing ${\cal H}$ in Eq. (27) with Eq. (5), we see that ${\cal H}=\rho$ is the energy density. One can verify the relations,
\begin{equation}
{\cal L}_{\cal F}{\cal H}_P=1,~~~~P{\cal H}_P^2={\cal F},~~~~{\cal L}=2P{\cal H}_P-{\cal H},
\label{28}
\end{equation}
where
\begin{equation}
{\cal H}_P=\frac{\partial {\cal H}}{\partial P}=-(1+2\beta {\cal F})^2.
\label{29}
\end{equation}
Equation (26) shows that the function ${\cal F}(P)$ is not a monotonic function. As a result, there is not a one-to-one correspondence between ${\cal F}$ frame  and $P$ frame \cite{Bronnikov} because ${\cal F}(P)$ is a multivalued function. Then there is not an exact electric-magnetic duality between these frames.
For weak fields, $\beta{\cal F}\ll 1$, both models (1) and (27) are converted to the Maxwell theory, ${\cal L}=-{\cal F}$.

\section{Asymptotic Reissner-Nordstr\"{o}m black holes}

 From Einstein's equation (13), we find the Ricci scalar:
\begin{equation}
R=\kappa{\cal T}.
\label{30}
\end{equation}
Replacing the trace of the energy-momentum tensor, Eq. (3), into (30) and using the relation ${\cal F}=-(1/2)[E(r)]^2$, one obtains the Ricci curvature
\begin{equation}
R=\frac{2\kappa\beta[E(r)]^4}{(1-\beta[E(r)]^2)^2}.
\label{31}
\end{equation}
At $r\rightarrow \infty$, the electric field, according to Eq. (22), goes to zero, $E(r)\rightarrow 0$, and therefore the Ricci scalar approaches zero, $R \rightarrow 0$. The spacetime at big distances from the black hole becomes Minkowski's spacetime.
By virtue of the asymptotic (22) we obtain from Eq. (31) the value of the Ricci scalar at $r\rightarrow\infty$:
\begin{equation}
R=2\kappa\beta\left(\frac{Q^4}{r^8}+\frac{4.74Q^{16/3}\beta^{2/3}}{r^{32/3}}+{\cal O}(r^{-12})\right).
\label{32}
\end{equation}
Thus, the Coulomb term gives the main contribution in the order of $r^{-8}$ to the Ricci scalar at $r\rightarrow\infty$.
The metric function entering a static, spherically symmetric line element in Eq. (15) and the mass function are given by
\begin{equation}
f(r)=1-\frac{2GM(r)}{r},~~~~M(r)=\int_0^r\rho(r)r^2dr=m-\int^\infty_r\rho(r)r^2dr,
\label{33}
\end{equation}
where $m$ is the mass of the black hole. With the help of the expression for the energy density (5) and Eqs. (18) and (19), we obtain the mass function
\begin{equation}
M(r)=m+\frac{ Q^{3/2}}{30\beta^{1/4}}y(-32+22xy-5x^2y^2),
\label{34}
\end{equation}
where $y$ is given by Eq. (20) with the notations (18) and the mass of the black hole is
\begin{equation}
m=M(\infty)=\frac{16 Q^{3/2}}{15\beta^{1/4}}.
\label{35}
\end{equation}
We have implied that the mass of the black hole possesses an electromagnetic nature.
The plot of the metric function $f(x)$ is represented in Fig. 3 for different values of the parameter $C=15\sqrt{\beta}/(GQ)=16\sqrt{Q}\beta^{1/4}/(Gm)$.
\begin{figure}[h]
\includegraphics[height=3.0in,width=3.0in]{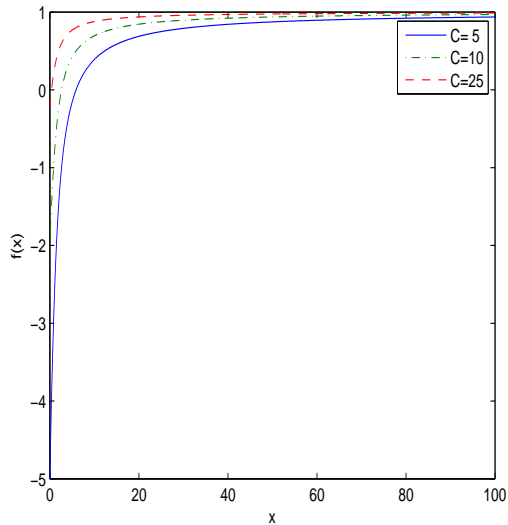}
\caption{\label{fig.3}The function $f(x)$ for $C=25, 10, 5$.}
\end{figure}
It follows that black holes have one (nonextreme) horizon for parameters $30\geq C\geq 0$ ($1\geq y\geq 0$).
The plot of the value of the horizon $x_+=r_+/(\beta^{1/4} \sqrt{Q})$ (when $f(r_+)=0$) vs $C$ is given in Fig. 4.
\begin{figure}[h]
\includegraphics[height=3.0in,width=3.0in]{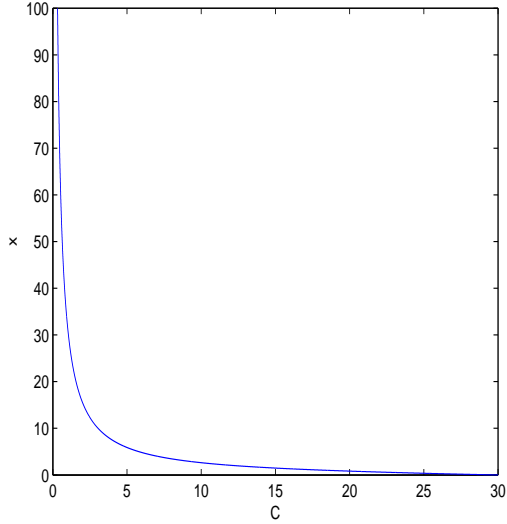}
\caption{\label{fig.4}The function $x_+$ vs $C$.}
\end{figure}
One can obtain the asymptotic value of the mass function (34) at $r\rightarrow \infty$ by the use of Eq. (21):
\begin{equation}
M(r)=m-\frac{ Q^{2}}{2r}-\frac{0.059Q^{10/3}\beta^{2/3}}{r^{11/3}}+\frac{0.021Q^{4}\beta}{r^{5}}+{\cal O}(r^{-19/3}).
\label{36}
\end{equation}
From Eqs.(33) and (36), we find the metric function
\begin{equation}
f(r)=1-\frac{2Gm}{r}+\frac{G Q^{2}}{r^2}+\frac{0.1185GQ^{10/3}\beta^{2/3}}{r^{14/3}}-\frac{0.042GQ^{4}\beta }{r^{6}}+{\cal O}(r^{-22/3}).
\label{37}
\end{equation}
The first three terms in Eq. (37) represent the RN solution and the last terms give corrections.
The spacetime at $r\rightarrow\infty$ asymptotically becomes the Minkowski spacetime. At $\beta=0$,  one comes to
Maxwell's electrodynamics, and solution (37) is converted to the RN solution.
There are different models of NLED \cite{Breton}, \cite{Hendi}, \cite{Kruglov6} that also give an asymptotic RN black hole solution with some corrections. It is seen from Eq. (37) that corrections to the RN solution change the event horizon. With the help of
Eqs. (24),(33) and (34), we obtain the regular asymptotic of the metric function at $r\rightarrow 0$:
\begin{equation}
f(r)=1-\frac{2 GQ}{\sqrt{\beta}}+\frac{\sqrt{Q}Gr}{\beta^{3/4}}- \frac{ Gr^2}{4\beta}+{\cal O}(r^{3}).
\label{38}
\end{equation}
It should be mentioned that the function $f(r)$ in Eq. (38) does not approach the value $1$ for $r \rightarrow 0$.
Therefore, the spacetime is not Minkowski's spacetime in the  neighborhood of the origin, and the regularity condition $f(r)\rightarrow 1$ for $r\rightarrow 0$ is violated. Thus, the solution obtained here is not a regular black hole in the usual sense. In addition, the strong energy condition should be broken for a regular black hole and the black hole must have at least
two horizons (or one degenerate horizon) \cite{Hawking1}. In our case, the equation $f(x)=0$ (see the plots in Fig. 3) possesses only one
root; i.e., the black hole has one horizon. According to the plots in Fig. 3, f(0) is negative, and, therefore, the metric is
nonstatic at the center of the black holes. I also mention that, at strong gravitational fields, quantum mechanics corrections must be taken into account. As a result, GR should break down and, hence, GR cannot be used to show a singularity \cite{Hawking2}.

\section{Thermodynamics}

In this section, we study the thermal stability of charged black holes. For this purpose one needs to calculate the temperature of the black hole and its heat capacity. The point where the temperature and  heat capacity change the sign corresponds to the first-order phase transition. The region of negative temperature corresponds to unstable state of the black hole. The second-order phase transition is related to the point where the heat capacity is singular.

To study the thermodynamics of black holes within our model, we calculate the Hawking temperature,
\begin{equation}
T_H=\frac{\kappa_S}{2\pi}=\frac{f'(r_+)}{4\pi},
\label{39}
\end{equation}
where $\kappa_S$ is the surface gravity and $r_+$ is the event horizon. From Eqs. (33), we obtain the following relations:
\begin{equation}
f'(r)=\frac{2 GM(r)}{r^2}-\frac{2GM'(r)}{r},~~~M'(r)=r^2\rho,~~~M(r_+)=\frac{r_+}{2G}.
\label{40}
\end{equation}
From Eqs. (39),(40), and (5), we find the the Hawking temperature expressed through the variables (18),
\begin{equation}
T_H=\frac{y_+\left[1-\frac{15}{C}y_+^2(1+y_+^4)\right]}{4\pi\sqrt{Q}\beta^{1/4}(1-y_+^4)},
\label{41}
\end{equation}
where the value $y_+$ is connected with the horizon $x_+$ by the relation $x_+=(1-y_+^4)/y_+$.
The parameter $C$ corresponds to the value of the horizon $x_+$ ($f(r_+)=0$) and is given by
\begin{equation}
C=\frac{32(1-y_+)+22x_+y_+^2-5x_+^2y_+^3}{x_+},
\label{42}
\end{equation}
The plot of the function $T_H4\pi\sqrt{Q}\beta^{1/4}$ vs $x_+=r_+/\sqrt{Q}\beta^{1/4}$ is given in Fig. 5.
\begin{figure}[h]
\includegraphics[height=3.0in,width=3.0in]{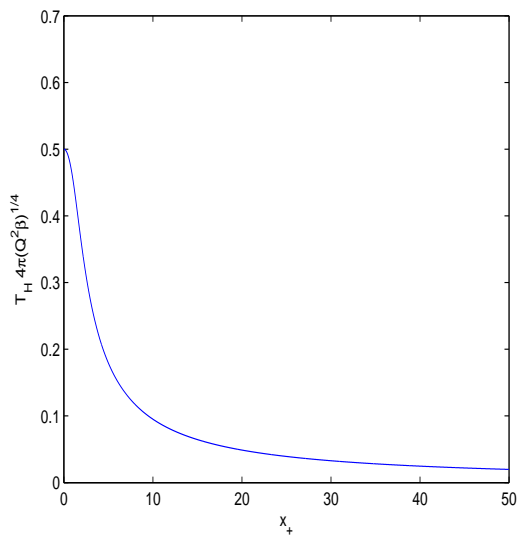}
\caption{\label{fig.5}The function $T_H4\pi\sqrt{Q}\beta^{1/4}$ vs $x_+$.}
\end{figure}
It follows from the graph that the temperature $T_H$ never becomes zero and, therefore, there is no phase transition of black holes
of the first-order.
Now we consider the heat capacity to study the possible phase transition of the second-order. The entropy satisfies the Hawking area low $S=A/4=\pi r_+^2$. The heat capacity at the constant charge is given by
\begin{equation}\label{43}
C_Q=T\frac{\partial S}{\partial T}|_Q.
\end{equation}
The calculation of  the heat capacity (43) gives the expression
\begin{equation}\label{44}
C_Q=\frac{-2\pi Q\sqrt{\beta}(1-y_+^4)^2 \left[1-\frac{15y_+^2}{C}(1+y_+^4)\right]}{y_+^2\left[1-\frac{15y_+^2}{C}(3-y_+^4)\right]},
\end{equation}
where the parameter $C$ obeys Eq. (42) and $x_+=(1-y_+^4)/y_+$.
The plot of the function $C_Q$ vs $x_+$ is given in Fig. 6.
\begin{figure}[h]
\includegraphics[height=3.0in,width=3.0in]{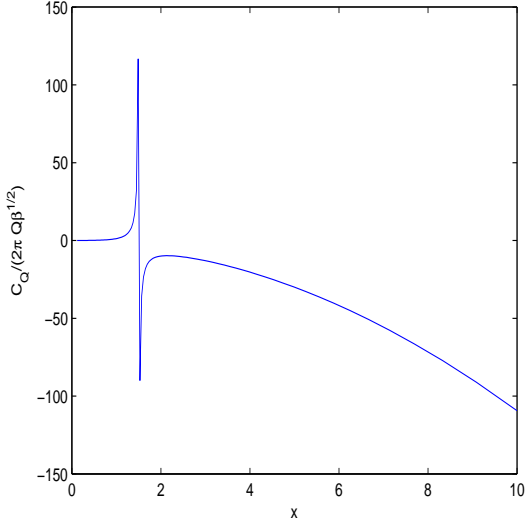}
\caption{\label{fig.6}The function $C_Q$ vs $x_+$.}
\end{figure}
The heat capacity $C_Q$ diverges (the denominator of $C_Q$ in Eq. (44) becomes zero) at the value of the horizon $x_+\simeq 1.506$ ($y_+\simeq 0.585822$). There is the second-order phase transition at this value of the horizon $x_+$ because the heat capacity becomes singular.
Then from Eq. (42) at this value $x_+\simeq 1.506$,  we obtain the constant $C\simeq 14.837$. It follows from Fig. 6 that when $x_+\geq 1.506$ ($r_+\geq 1.506 \sqrt{Q}\beta^{1/4}$), the heat capacity is negative and, therefore, the black hole is unstable. We find critical values of the parameter $\beta$, the mass of the black hole $m$, and Hawking's temperature corresponding to this horizon $x_+= 1.506$,
\[
\beta=\left(\frac{CQG}{15}\right)\simeq 0.98 Q^2G^2,~~~m=\frac{16 Q^{3/2}}{15\beta^{1/4}}\simeq 1.07\frac{Q}{\sqrt{G}},
\]
\begin{equation}\label{45}
T_H\approx \frac{1}{40\pi Q\sqrt{G}}.
\end{equation}
Thus, within the NLED considered, if the black hole mass is greater than the Planck mass (times the charge of the black hole), the black hole undergoes the second-order phase transition and becomes unstable.

\section{Conclusion}

In the model of NLED proposed, we studied energy conditions and showed that WEC, DEC, and SEC are satisfied.
For the case of spherical symmetry we have obtained the exact solution for the electric field of charged objects and found
the corrections to the Coulomb law that are in the order of $r^{-14/3}$. It was demonstrated that the electric field and the electric
potential are finite at the center of the charged objects and there are no singularities.
The P framework and the electric-magnetic duality transformations were considered.
NLED coupled to the gravitational field was investigated, and we calculated the Ricci scalar and its asymptotic at $r\rightarrow \infty$ which is in the order of $r^{-8}$. We obtained the mass of the black hole expressed through the parameter $\beta$ and the charge.
The asymptotic of the metric function at $r\rightarrow\infty$ was found, and corrections to the Reissner-Nordstr\"om solution were obtained.
We found the metric function at $r\rightarrow 0$. The thermodynamics of black holes and phase transitions were investigated. We showed tha,t in our model, there is no phase transition of the first-order because the temperature of the black hole does not change the sign. But at the critical value of the horizon, the phase transition of the second-order
takes place and the heat capacity is singular. If the mass of the black hole is greater than the critical value, the black hole becomes unstable.

\end{document}